\documentstyle[aps,prl,multicol,epsf]{revtex}

\begin{document}
\draft

\title{Interface Depinning in the Absence of External Driving Force} 

\author{Jos\'e J. Ramasco$^{1,2,}$\cite{mail1}, 
Juan M. L\'opez$^{3,1}$\cite{mail2} and
Miguel A. Rodr\'{\i}guez$^{1}$
}

\address{$^1$ Instituto de F\'{\i}sica de Cantabria, 
CSIC-UC, E-39005 Santander, Spain}  

\address{$^2$ Departamento de F\'{\i}sica Moderna,   
Universidad de Cantabria, E-39005 Santander, Spain}   

\address{$^3$ INFM Sezione di Roma 1, 
Universit\`a di Roma "La Sapienza", 
P.le Aldo Moro 2,
I-00185 Roma, Italy}

\maketitle

\begin{abstract}
We study the pinning-depinning phase transition 
of interfaces in the quenched Kardar-Parisi-Zhang model 
as the external driving force $F$ goes towards zero. 
For a fixed value of the driving force we induce depinning 
by increasing the nonlinear term coefficient $\lambda$, 
which is related to lateral growth, up to a critical threshold. 
We focus on the case in which there is no external force 
applied ($F=0$) and find that, contrary to a simple scaling 
prediction, there 
is a finite value of $\lambda$ that makes the interface to 
become depinned. The critical exponents at the transition are
consistent with directed percolation depinning. Our results 
are relevant for paper wetting experiments, in which an interface
gets moving with no external driving force.    

\end{abstract}


\begin{multicols}{2}
\narrowtext

The dynamics of random interfaces in the presence of noise
is an interesting example of critical phenomena and
generic scale-free behaviour in systems 
far from equilibrium.
In the case of surface growth dominated by 
thermal fluctuations,
the Kardar-Parisi-Zhang (KPZ) equation \cite{kpz} 
has been very much studied for it represents a 
whole universality class
of growth, which includes many well-known discrete computer 
models \cite{barabasi}. In many experimental 
situations, however, interface motion is affected 
by the existence of random pinning forces (see \cite{barabasi} 
and references therein). 
In this case, the simplest way to model interface 
roughening is to replace the noise term 
$\eta({\bf x},t)$ in KPZ
by a quenched disorder $\eta({\bf x},h)$, 
\begin{equation}
\label{qkpz}
{\partial h \over \partial t} = \nu {\bf \nabla}^2 h + 
\lambda (\nabla h)^2 + F + \eta({\bf x},h),
\end{equation}
which is often referred to as the quenched 
Kardar-Parisi-Zhang (QKPZ) 
equation. The first term on the right-hand side 
describes the smoothening effect of surface tension, 
$F$ is the driving force that pushes 
the interface through 
the disorder, and the term $\lambda (\nabla h)^2$
comes from lateral growth and represents 
the nonlinear most relevant correction.
The quenched disorder has 
short-range correlations
$\langle \eta({\bf x},h) \eta({\bf x'},h')\rangle = 
\delta({\bf x}-{\bf x}') \Delta(h-h')$, 
where the correlator
$\Delta(u)$ is a very rapidly decreasing function 
of $|u|$ and is
the term actually responsible for the pinning 
of the interface. This equation is expected to describe
interface roughening in
many disordered systems, including
the non-equilibrium dynamics 
of magnetic domain walls in 
disordered materials \cite{bruinsma,fisher,narayan,jost}, 
an elastic chain in a
quenched disorder \cite{henrik}, fracture cracks 
propagation \cite{ertas}, 
{\it etc}). Its applicability to describing 
fluid-fluid displacement in porous
media might be less justified though \cite{dube}.

The QKPZ model described by (\ref{qkpz}) 
exhibits a continuous phase transition at 
a certain critical value, $F_c$, 
of the external driving force $F$. For $F$ larger than
$F_c$, the interface moves with a finite 
velocity. However, the interface remains pinned 
by the disorder for $F<F_c$. The critical
point $F=F_c$ is known as depinning transition. 
The interface velocity scales as 
$v \sim (F-F_c)^\theta$ near and
above the transition and plays the role 
of an order parameter.

The value of the 
critical force depends on the parameters of the model, 
in particular depends on the value of
the coefficient $\lambda$ of the nonlinear term. 
Therefore, by keeping constant the rest of the 
equation parameters one can 
find a critical line $F_c=f(\lambda)$ separating 
the pinned from the depinned phase. Alternatively, 
we can see this critical
line the other way around and let 
$\lambda_c=f^{-1}(F)$ be the 
critical value of the KPZ nonlinearity above 
which the interface gets depinned. The driving 
force $F$ favors the advance of the interface
and thus, the lower the driving force is, the larger 
the critical value $\lambda_c$ of the nonlinearity 
that is needed in order
to get the interface depinned. Indeed one would 
expect that as $F \to 0$ depinning 
becomes more and more difficult until eventually, 
at $F=0$, the threshold $\lambda_c \to \infty$ and 
depinning becomes impossible. This 
intuitive picture can be justified by means of a 
simple scaling argument
as follows. Consider a typical region of size 
$l$ pinned by the disorder.
Eq.(\ref{qkpz}) applied to that region reads 
\begin{equation}
\label{balance}
\nu h l^{-2} + \lambda h^2 l^{-2} + F - 
\Delta(0)^{1/2} l^{-d/2} = 0.
\end{equation}
If one supposes that the nonlinear term dominates over 
the diffusion, the interface remains pinned whenever 
$\lambda a^2 l^{-2} \ll \Delta(0)^{1/2} l^{-d/2}$, 
where $a$ is the lattice spacing in the growth direction. 
This defines a characteristic length, 
$l_c = [\lambda^2 a^4 / \Delta(0)]^{1/(4-d)}$, 
such that for $l \ll l_c$ the interface gets pinned. 
Now to estimate the critical force that 
is necessary to depin a region of typical size $l_c$, 
one equates the force term with the disorder 
in Eq.(\ref{balance}) to get
to an expression for the critical line,
$F_c \sim \Delta(0)^{2/(4-d)} (\lambda a^2)^{-d/(4-d)}$. 
Inverting the latter, one finds 
\begin{equation}
\label{cline}
\lambda_c \sim { \Delta(0)^{2/d} \over a^2} F^{-(4-d)/d}
\end{equation}
for the critical line of the depinning transition \cite{linear}. 
In $1+1$ dimensions for instance, 
Eq.(\ref{cline}) predicts a diverging 
$\lambda_c \sim F^{-3}$ as $F \to 0$.

In this Letter, we show that, contrary to  
this scaling picture, there is always a finite 
critical value $\lambda_c$ of the KPZ nonlinearity 
such that the interface gets depinned even for $F=0$. 
Our conclusions are based upon 
numerical integration of Eq.(\ref{qkpz}) in $d=1$.
We numerically calculate the critical line and find 
that $\lambda_c (F=0) = 3.60 \pm 0.01$
(in natural units) for the QKPZ 
equation. Our results support the somehow 
counterintuitive conclusion that an interface 
can get depinned in absence 
of external driving force by the solely 
effect of nonlinearities. 

In order to numerically integrate Eq.(\ref{qkpz}),  
the equation parameters
can easily be rescaled to have only two independent 
tuning parameters-- namely, the nonlinear
KPZ coefficient $\lambda$ and the driving force $F$. 
We have used a standard finite-differences 
scheme for integrating 
the QKPZ equation given (in natural units) by
\begin{eqnarray}
\label{discrete}
h(i,t+\Delta t) & = & h(i,t) + \Delta t \:\:\: F 
+ \Delta t \:\:\: \eta[i,\tilde h(i,t)]\\
\nonumber
& + & \Delta t \:\:\: [h(i+1,t)+h(i-1,t)-2h(i,t)] \\
\nonumber
& + & \Delta t \:\:\: \lambda [{h(i+1,t) 
- h(i-1,t) \over 2}]^2,
\end{eqnarray}
where the lattice spacing has been set to unity.
We start our simulation from a flat initial
condition $h(x,0) = 0$ and
periodic boundary conditions, 
{\it i.e.} $h(0,t) = h(L,t)$ and $h(L+1,t) = h(1,t)$,
are imposed on the interface. $\tilde h(i,t)$ stands 
for the integer part 
of $h(i,t)$, and the quenched disorder is Gaussian 
distributed and has correlations 
$\langle \eta(i,\tilde h) \eta(j,\tilde h') \rangle = 
\delta_{i,j} \delta_{\tilde h,\tilde h'}$. 
Simulations with different
time steps were carried out, 
and the scheme proved to be stable and well behaved 
for a time step $\Delta = 0.01$ (or smaller) for the 
range of tuning parameters simulated.

We carried out simulations in systems of 
size $L=128, 256, \cdots, 8192$. For each
value of the of the nonlinear coefficient $\lambda$
we computed the critical value of force 
needed to get the interface depinned. 
Our results are summarized 
in Fig. 1. As expected, we find that as the
driving force is smaller the critical value 
$\lambda_c$ of 
the nonlinear coefficient required
in order to depin the interface becomes larger. 
However, as anticipated above, the critical
point $\lambda_c$ always remains finite, 
even for $F =0$. At a purely phenomenological level, 
we find that the critical line can  
be fitted very nicely by 
\begin{equation}
({\lambda \over b_1})^{2/3} + ({F \over b_2})^{2/3} = 1,
\end{equation}
where the constants $b_1 = 4.31 \pm 0.04 $ and 
$b_2 = 0.81 \pm 0.03$ (see Fig. 1). To our knowledge 
this is the first formula for the critical line 
and demands theoretical explanation.

In the following we focus on the case in which 
no external driving, $F=0$, pushes the interface 
and depinning is due solely to nonlinear lateral growth. 
We have studied the critical behaviour in the 
vicinity of $\lambda_c(F=0) = 3.60 \pm 0.01$ in order 
to address the problem of the nature of the critical point. 
Firstly, we have computed the scaling behaviour of the 
stationary interface velocity at $F=0$ as the transition 
is approached. In Figure 2 (inset) 
we plot $v$ {\it vs.} $\lambda$ 
for $F=0$ and a system of size $L=8192$ showing that 
the transition is continuous. The critical behaviour 
of the order parameter $v$ is shown in Figure 2. 
We find that close to the depinning threshold the 
interface velocity scales as 
$v \sim (\lambda -\lambda_c)^\theta$ with a critical 
exponent $\theta = 0.635 \pm 0.007$.

The depinning mechanism for $F=0$ is the following.
Starting from a flat initial condition $h(x,t=0)=0$ all
the terms in Eq.(\ref{qkpz}) are zero except for the disorder.
At time $t=0$ the quenched random term $\eta(x,h)$ 
generates inhomogeneities in
the front, which in turn produce a finite value of 
$(\nabla h)^2$. For small values of $\lambda$ this
inhomogeneities smear out and the interface gets pinned by
the disorder at one of the infinite pinning paths. However,
for $\lambda > \lambda_c$ these initial inhomogeneities
are effectively amplified by the nonlinearity
and the interface gets moving with a finite velocity.

As occurs in the standard case 
of depinning at a threshold value
of the driving force $F=F_c$, 
the depinned phase is rough and 
belongs to the universality class of 
KPZ. This can be seen by studying 
the scaling behaviour of the the global width 
$W(L,t) = [\langle h(x,t)^2 \rangle - 
\langle h(x,t) \rangle^2]^{1/2}$, 
where the average is over all $x$ and 
different realizations
of disorder \cite{anom}. We obtain that 
the global width scales as 
\begin{equation}
\label{W-fv}
W(L,t) \sim
\left\{ \begin{array}{lcl}
     t^\beta & {\rm if} & t \ll t_{\times}\\
     L^{\alpha}     & {\rm if} & t \gg t_{\times}
\end{array}
\right. ,
\end{equation}
with a time exponent 
$\beta = 0.33 \pm 0.01$ and a roughness exponent 
$\alpha = 0.50 \pm 0.01$ in agreement
with the KPZ class of growth. 

However, when approaching the depinning transition 
from above, 
$\epsilon = (\lambda - \lambda_c)/ \lambda_c \to 0^{+}$, 
the scaling of the global width is 
affected by the existence of a diverging
correlation length 
$\xi \sim \epsilon^{-\nu}$. 
This is the typical size of the fluctuations 
of the majority phase, {\it i.e.} the 
characteristic size of connected regions formed by 
pinned sites. As we show in Figure 3, the global width
(and similarly does the local width)
displays a crossover from $\sim t^{0.7}$ to 
KPZ-like behaviour $\sim t^{0.33}$. 
More precisely, one can see in Fig. 3 that 
the width approximately behaves as
\begin{equation}
\label{w_lambda}
W(t,\epsilon) \sim 
\left\{ \begin{array}{lcl}
     t^{\beta_c} \epsilon^{\kappa_c} & {\rm if} & t \ll t_c\\
     t^{\beta_{kpz}} \epsilon^{-\kappa} & {\rm if} & t \gg t_c
\end{array}
\right. ,
\end{equation}
where $\kappa_c$, in view of the dependence of 
the curves on $\epsilon$, must be very small. 
These two regimes
are separated by a crossover time $t_c$ that
depends on $\epsilon$.
Indeed, following Kertesz and Wolf 
\cite{kertesz&wolf}, near a roughening phase transition 
one expects the crossover time to scale
with the distance to the threshold as 
$t_c \sim \xi^z \sim \epsilon^{-\gamma}$, where $\gamma = z \nu$. 
Direct examination of Figure 3
immediately suggests the scaling ansatz 
\begin{equation}
\label{rt}
W(t,\epsilon) \sim t^{\beta_{kpz}} \epsilon^{-\kappa} g(t/t_c),
\end{equation}
which is characteristic of systems close to 
a roughening transition 
\cite{kertesz&wolf,makse&amaral,lopez&jensen}.
The scaling function is given by
\begin{equation}
\label{rt-f}
g(u) \sim 
\left\{ \begin{array}{lcl}
     u^{\beta_{c} - \beta_{kpz}} & {\rm if} & u \ll 1\\
     {\rm const.} & {\rm if} & u \gg 1
\end{array}
\right. ,
\end{equation}
and the scaling relation 
\begin{equation}
\label{scal-rel}
\kappa_c + \kappa = (\beta_c - \beta_{kpz}) \gamma  
\end{equation}
among critical exponents must be fulfilled so that 
both regimes match. 

In Figure 3 (inset) we show a
data collapse of $t^{-\beta_{kpz}}\epsilon^{\kappa}W(t,\epsilon)$ 
{\it vs.} $\epsilon^{\gamma}t$. A good data collapse is obtained for 
the exponents $\beta_{kpz} = 0.3$, $\kappa = 0.57$ and $\gamma = 1.57$,
the error in estimating these exponents being of about 
10{\%}. From the scaling relation (\ref{scal-rel}) one 
also gets $\beta_c = 0.73$ in good
agreement with our previous estimate.

The value of the critical exponents is consistent with those 
of the DPD model \cite{tang&lesch,buldyrev} just above the 
transition \cite{makse&amaral,barabasi}.
We thus conclude that the lateral growth driven 
depinning point at $F=0$ and $\lambda=\lambda_c$ also 
belongs to the universality class of DPD. 

Our results indicate that in the absence of any 
external driving field an interface can get depinned by 
increasing the nonlinear term $\lambda$ in (\ref{qkpz}) up to
its critical value. From the experimental point of view, 
this implies that, assuming the parameter $\lambda$ 
is tunable in the laboratory, an interface
could become depinned even with when 
no external driving force is applied. 
In the following we discuss the role of 
anisotropy of the background random medium in
generating the KPZ term $\lambda (\nabla h)^2$, and how 
this mechanism can be used to rise the value of 
$\lambda$ in experiments by increasing the
degree of disorder anisotropy.

The QKPZ equation for $\lambda=0$ is known as the quenched
Edwards-Wilkinson (QEW) equation and has been much studied
in recent years. The critical exponents at the depinning transition
have been well determined by several 
authors \cite{amaral,leschhorn,roux,lopez&rodriguez,henrik}. In 1+1 
dimensions one finds $\alpha \sim 1.25$ and $\beta \sim 0.85$ 
at the threshold $F=F_c$ and $\alpha = 1/2$ and $\beta = 1/4$ 
in the moving phase for $F \gg F_c$, 
where the disorder $\eta(x,h)$
can be replaced by $\eta(x,vt)$ and the exponents of the 
EW universality class \cite{barabasi} are recovered. 
The QEW equation arises naturally as the Langevin equation for the 
Hamiltonian $H = \int d{\bf x} [\sqrt{1+(\nabla h)^2} 
+ V({\bf x},h)]$ describing the elastic 
energy of an interface in a disordered potential 
$V({\bf x},y)$ \cite{bruinsma,fisher,narayan}.  
The term $\lambda (\nabla h)^2$ cannot be deduced as a 
variation of any Hamiltonian and is
added as the most relevant nonlinear 
correction \cite{barabasi}. Geometrically, it 
accounts for growth in a direction
locally normal to the interface and is referred to as 
nonlinear lateral growth term. 

In the past the physical origin of the KPZ 
nonlinearity in interface depinning has been 
found to be related to two distinct mechanisms 
for different models \cite{stepanow}. 
On the one hand, in the spirit of the original work 
of KPZ \cite {kpz}, the $\lambda$ term can have a
purely kinematic origin, so that 
$\lambda \propto v$ \cite{amaral,stepanow}. 
In this case, the term $\lambda(\nabla h)^2$ goes 
to zero at the depinning transition, $F=F_c$, and
the system thus belongs to the QEW universality class.   
On the other hand, there are models \cite{stepanow} 
for which $\lambda$ remains finite at the 
transition \cite{finite}. These 
models have exponents that correspond to the DPD 
universality class \cite{amaral,neshkov}. 
Tang, Kardar and Dhar \cite{tang} have shown that
this finite $\lambda$ term can arise in some models 
because of an underlying anisotropy in the 
random medium, {\it i.e.} models that have a
growth direction determined by the random medium.
A further numerical step on this direction 
has recently been achieved by Park, Kim and 
Kim \cite{park} by studying a model with an anisotropic 
disorder correlator. 
The effect of anisotropy on real experiments
has also been successfully tested 
by Albert {\it et. al.} \cite{albert}.
Experiments on fluid flow in a random medium formed
by packed glass beads \cite{rubio} are now known to belong
to the isotropic QEW universality class \cite{albert}.
However, the scaling exponents obtained for paper wetting 
\cite{buldyrev,horvath} are in excellent agreement with 
the prediction of the anisotropic DPD universality class.
In paper wetting experiments a sheet of paper 
is vertically suspended over a reservoir of liquid (usually
black ink). The fluid then wets the paper and the interface 
between wet and dry phases rises until it eventually
stops. The interface grows upwards because of capillarity
forces in the paper pores. Notice that there is
none external driving force. The anisotropic paper 
fibre distribution determines the local capillarity 
forces. Disorder in these systems is thus highly anisotropic.
We believe that this system is an excellent example of
depinning driven solely by the nonlinear lateral growth term. 

In summary, we have studied the QKPZ equation focusing on
the case in which there is no external driving force ($F=0$).
We have shown that there exists a depinning transition for a 
finite value of the KPZ coefficient $\lambda = \lambda_c(F=0)$
and that transition belongs to the DPD universality class. 
A finite value of the 
nonlinear coefficient $\lambda$ appears in 
systems with anisotropic disorder like for instance in 
paper wetting experiments. In this system there is 
no external driving force and depinning 
occurs due to local capillarity forces, 
which drive the interface
through the anisotropic lateral growth term $\lambda(\nabla h)^2$.  
We conclude that by varying the anisotropy degree 
of the corresponding random medium
in other experimental systems, depinning is possible even with no
external driving. 

We thank S. Zapperi and S. Stepanow for discussions. 
Financial support from
DGES of the Spanish Government (Project No. PB96-0378-C02-02) is
acknowledged. J.\ J.\ R. is supported by a FPI fellowship of the 
Ministerio de Educaci\'on y Cultura (Spain). J.\ M.\ L. 
was supported by a TMR Network of the European 
Commission (Contract No. FMRXCT980183) at INFM 
and a Marie Curie Return Fellowship at IFCA.

\begin{figure}
\centerline{
\epsfxsize=7.5cm
\epsfbox{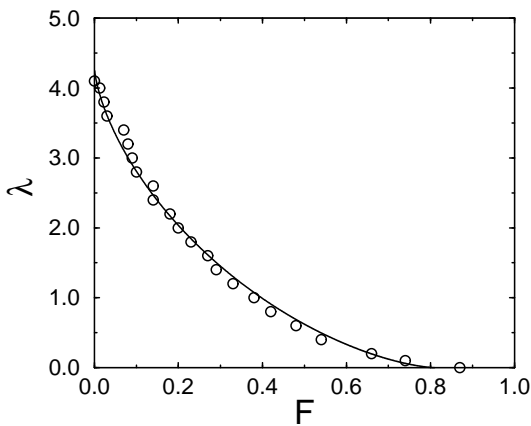}}
\caption{Critical line $\lambda_c=f(F)$ for the 
QKPZ equation. Symbols are points obtained 
from numerical simulations in a system of size $L=1024$.
The line is a fit according to Eq.(5). Note that 
$\lambda_c$ remains finite, even at $F=0$.}
\end{figure}

\begin{figure}
\centerline{
\epsfxsize=7.5cm
\epsfbox{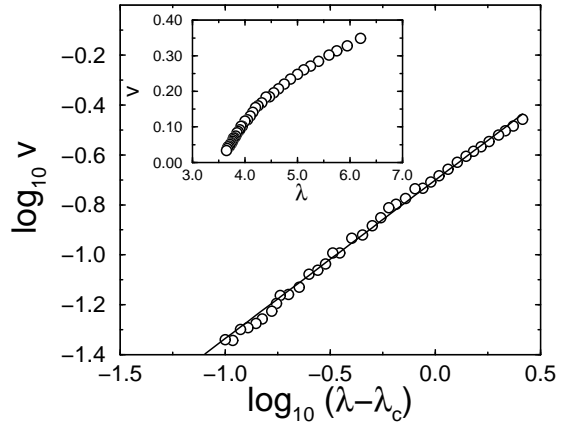}}
\caption{Interface velocity {\it vs.} coefficient $\lambda$ for 
the QKPZ equation at $F=0$ (inset) close to the threshold
$\lambda_c(F=0)$. The critical behaviour of the velocity
$v \sim (\lambda-\lambda_c)^\nu$ is shown in 
the main panel. A straight line is found for 
$\lambda_c=3.60 \pm 0.01$ and the slope corresponds
to the velocity critical exponent $\nu = 0.635 \pm 0.007$.}
\end{figure}

\begin{figure}
\centerline{
\epsfxsize=7.5cm
\epsfbox{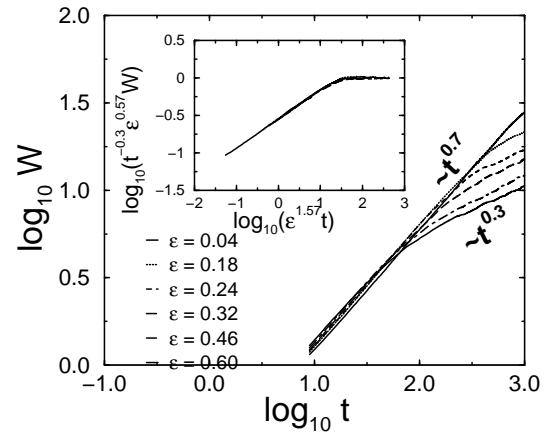}}
\caption{In main panel we plot the global width  
for different distances (as shown)
$\epsilon = (\lambda - \lambda_c)/\lambda_c$ to the 
threshold for $F=0$ in a system of size $L=8192$. The crossover
from $t^{0.7}$ to $t^{0.3}$ occurs at times that scale as
$t_c \sim \epsilon^{-\gamma}$ with the distance to the threshold.
Inset shows a data collapse according to Eq.(8) 
of the sets shown in the main panel. A good collapse is found for
the exponents $\beta_{kpz}= 0.3$, $\kappa = 0.57$ 
and $\gamma = 1.57$.} 
\end{figure}

\end{multicols}

\begin{thebibliography}{99}
\bibitem{linear} In the case of depinning for $\lambda = 0$, 
the diffusion term must equilibrate the disorder and one gets to
a driving force $F_c = [\Delta (0)^{4/d}/(\nu a)]^{d/2(4-d)}$ 
instead of Eq.(\ref{cline}).
\bibitem{anom} Following \cite{jose}, we also 
checked that there is no anomalous scaling 
phenomena by computing the local width and power spectrum.
\bibitem{finite} Notice that, although it was first though
that in DPD models $\lambda \to \infty$ as the depinning
transition is approached, it has recently been 
shown \cite{neshkov} that
this inconsistency was due to the incorrect definition of the 
tilt-velocity dependence used in \cite{amaral}. In fact, 
$\lambda$ remains finite at the transition \cite{neshkov}.
\end{thebibliography}
\end{document}